\RequirePackage{snapshot}
\documentclass[fleqn,usenatbib]{mnras}

\usepackage{newtxtext,newtxmath}

\usepackage[T1]{fontenc}

\DeclareRobustCommand{\VAN}[3]{#2}
\let\VANthebibliography\thebibliography
\def\thebibliography{\DeclareRobustCommand{\VAN}[3]{##3}\VANthebibliography}

\usepackage{graphicx}	
\usepackage{amsmath}	

\newcommand{\Msun}{M$_{\odot}$} 
\newcommand{\microgauss}{$\mu$G} 

\newcommand{\aref}[1]{\hyperref[#1]{Appendix~\ref{#1}}}

\usepackage{color}
\usepackage[normalem]{ulem}
\definecolor{darkgreen}{rgb}{0.13, 0.55, 0.13}

\title[Simulated Milky Way magnetic fields]{The global structure of magnetic fields and gas in simulated Milky Way-analogue galaxies}

\author[B. D. Wibking et al.]{
Benjamin D. Wibking,$^{1}$\thanks{E-mail: ben.wibking@anu.edu.au (BDW)}
and Mark R. Krumholz$^{1}$
\\
$^{1}$Research School of Astronomy \& Astrophysics, Mount Stromlo Observatory, Cotter Road, Weston Creek, ACT 2611 Australia\\
}

\date{Accepted XXX. Received YYY; in original form ZZZ}

\pubyear{2021}

\begin{document}
\label{firstpage}
\pagerange{\pageref{firstpage}--\pageref{lastpage}}
\maketitle

\begin{abstract}
    We simulate an isolated, magnetised Milky Way-like disc galaxy using a self-consistent model of unresolved star formation and feedback, evolving the system until it reaches statistical steady state. We show that the quasi-steady-state structure is distinctly layered in galactocentric height $z$, with an innermost region having comparable gas and magnetic pressures (plasma beta $\beta \sim 1$), an outermost region having dominant gas pressures ($\beta \gg 1$), and an intermediate region between $300$ pc $\lesssim |z| \lesssim 3$ kpc that is dynamically dominated by magnetic fields ($\beta \ll 1$). We find field strengths, gas surface densities, and star formation rates that agree well with those observed both in the Galactic centre and in the Solar neighbourhood. The most significant dynamical effect of magnetic fields on the global properties of the disc is a reduction of the star formation rate by a factor of 1.5--2 with respect to an unmagnetised control simulation. At fixed star formation rate, there is no significant difference in the mass outflow rates or profiles between the magnetised and non-magnetised simulations. Our results for the global structure of the magnetic field have significant implications for models of cosmic ray-driven winds and cosmic-ray propagation in the Galaxy, and can be tested against observations with the forthcoming Square Kilometre Array and other facilities. Finally, we report the discovery of a physical error in the implementation of neutral gas heating and cooling in the popular \textsc{gizmo} code, which may lead to qualitatively incorrect phase structures if not corrected.
\end{abstract}




\section{Introduction}

The role of magnetic fields in the Galaxy has been long debated. The discovery of polarised starlight by \cite{Hiltner_1949} and \cite{Hall_1949} led to the hypothesis that the observed polarisation was caused by dust grains aligned with the magnetic field in the interstellar medium, with the immediate implication from the observed polarisation direction that Galactic magnetic fields are aligned parallel to the Galactic plane, i.e. are toroidal \citep{Davis_1949,Davis_1951}. At nearly the same time, \cite{Fermi_1949} theorised that fluctuating magnetic fields in the interstellar medium were the origin of cosmic rays (via `second-order Fermi acceleration'). Soon afterward, the Galactic radio emission first observed by Jansky and Reber was proposed to be due to the radiation produced by the gyration of cosmic rays around magnetic fields \citep{Kiepenheuer_1950}. These theoretical advances were followed by radio observations of polarised Galactic emission and Faraday rotation in the interstellar medium in the 1950s and 1960s (see \citealt{Wielebinski_2012} for a detailed historical review of polarised radio observations).

In parallel with this observational progress, theorists began to consider the possible dynamical role of magnetic fields. This topic was first considered by \cite{Alfven_1942}, who argued that magnetic fields may be dynamically important on the surface of the sun through their wave interactions with partially-ionised gas, and \cite{Fermi_1949}, who recognized that the same considerations applied to the interstellar medium, further proposed that the energy of the Galactic magnetic field should be of order the energy of the turbulent motions of the gas, implying a magnetic field strength of order $\sim$ \microgauss. However, the topology of the field, and in particular its structure as one moves away from the Galactic plane, remain uncertain. One possible picture is provided by \cite{Parker_1966}, who discovered a magnetohydrodynamic instability of toroidal fields that leads to the magnetic field buckling into loops above the plane. However, Parker's instability was suppressed in the magnetohydrostatic model of \cite{Boulares_1990}, which proposed a significant vertical (i.e., poloidal) component of the magnetic field at $\sim$ kpc heights above the Galactic plane and allowed for the diffusion of cosmic rays within a region of tangled (but mean-toroidal) magnetic fields near the plane, yielding a solution with approximate energy equipartition between turbulent motions, cosmic rays, and magnetic fields, and an extended distribution of H~\textsc{ii} gas at heights $|z| \gtrsim 1$ kpc (the so-called `Reynolds layer'; \citealt{Reynolds_1989}). This basic picture of the magnetohydrostatic steady-state structure of the gas, cosmic rays, and magnetic fields of the Galaxy has survived to the present.

The uncertainty in the field topology is only one of several outstanding questions about the origin and dynamical role of the magnetic field. One major area of uncertainty is the origin of the field, and whether it is governed by a mean field dynamo process \citep[e.g.,][]{Gent_2021}. A second is how magnetic fields interact with galactic winds driven by either supernova breakout \citep[e.g.,][]{Tomisaka_1998} or by cosmic rays \citep[e.g.,][]{Breitschwerdt_1991,Everett_2008,Mao_2018,Quataert_2021}. There have been some attempts to address these questions in magnetohydrodynamic (MHD) cosmological simulations (e.g., \citealt{Pakmor_2013, Pakmor_2017, Pakmor_2018, Hopkins_2020}; we note that \citealt{Pakmor_2013} carried out isolated simulations but with comparable resolution and physics to cosmological simulations), but these efforts have been somewhat limited by resolution. Non-zoom-in cosmological simulations, such as the highest-resolution Illustris-TNG simulations \citep{Pillepich_2019,Nelson_2019}, have a resolution of $\Delta x \sim 150$ pc at the mean density of the Milky Way's interstellar medium ($n_{\rm H} \sim 1$ cm$^{-3}$), meaning the gas scale height of the Milky Way ($h\sim 100$ pc) is resolved with less than one gas particle in the diffuse interstellar medium. Zoom-in cosmological simulations do only slightly better, reaching $\Delta x \sim 60$ pc at $n_{\rm H} \sim 1$ cm$^{-3}$, thereby resolving the scale height with $\lesssim 2$ gas particles \citep[e.g.,][]{Pakmor_2017, Hopkins_2020}. This is clearly insufficient to capture the topology of the field and its changes in and out of the plane.

On the other end of the resolution spectrum lie MHD simulations of local patches of the Milky Way, such as those of \citet{Kim_2016,Kim_2019,Kim_2020}, \citet{Kim_2017,Kim_2018}, and \citet{Rathjen_2021}. These simulations benefit from uniform parsec-scale resolution of all of the phases of the interstellar medium, but they cannot resolve the global structure of the disc in radius and height, which means that cannot address questions of the field topology. Additionally, as emphasised by \cite{Martizzi_2016}, any local simulation does not allow streamlines to diverge in outflows that would otherwise be spherical, thus preventing the outflow from crossing the sonic point of a classical hot superwind \citep{Chevalier_1985}. This limits their ability to study the interaction of magnetic fields with outflows.

These limitations in previous work motivate us to consider an intermediate resolution regime. We present a new dynamical model of the Galaxy, evolved for $\sim 1$ Gyr until it has reached a quasi-steady-state. We reach roughly an order of magnitude higher mass resolution than even zoom-in cosmological simulations of Milky Way-like galaxies, and, though our resolution is still substantially smaller than that in local patch simulations, we retain the full geometry of the problem, so that we can study field topology and outflows. Relatively few simulations of this type have been published, and those have largely been concerned with studying the magnetisation of the neutral medium \citep[e.g.,][]{Wang_2009} or attempting to explain and interpret the observed Faraday rotation sky \citep[e.g.,][]{KulpaDybel_2015,Butsky_2017}. There have been no previous efforts to use simulations of this type to map out the vertical structure and topology of the magnetic field, or to study how fields interact with galactic winds. These questions are the focus of our study.

This paper is organised as follows: \autoref{section:methods} details the numerical methods used in our simulations, including the `subgrid models' used for star formation and feedback; \autoref{section:simulations} describes the initial conditions and evolution of our simulations as they relax into a quasi-steady-state; \autoref{section:discussion} summarises the zonal structure of the simulations in steady state, the dependence of outflow rate on star formation rate, and the observations to which our simulations may be compared. We conclude in \autoref{section:conclusions} with a summary of our results and possible future directions for research.

\section{Methods}
\label{section:methods}
We solve the equations of ideal magnetohydrodynamics (MHD) using the \textsc{GIZMO} code \citep{Hopkins_2015}, which implements the method of \cite{Hopkins_2016} and \citet{Hopkins_2016b}. This method significantly improves upon the divergence-cleaning approach of \cite{Dedner_2002} by the addition of a local approximate Hodge projection that further suppresses numerical magnetic monopoles in the discrete magnetic field. Note that we use the MHD solver even when running simulations with zero magnetic field, in order to ensure that our results are not biased by the use of a different solver in different simulations.

In addition to MHD, we solve a time-dependent chemistry network for the abundance of H~\textsc{i}, H~\textsc{ii}, He~\textsc{i}, He~\textsc{ii}, He~\textsc{iii}, and free electrons that we use to compute the atomic cooling rates for hydrogen and helium. Additionally, we interpolate from a table of metal line cooling rates as a function of density and temperature (assuming ionization equilibrium and solar metallicity) that was computed using Cloudy \citep{Ferland_1998} following the method of \cite{Smith_2008}, as implemented in the \textsc{Grackle} chemistry and cooling library \citep{Smith_2017}.\footnote{This split treatment of the ionization state is a reasonable approximation because almost all free electrons in the ISM come from the ionization of hydrogen and helium. Ionization equilibrium cannot be assumed to hold in general because the warm neutral medium has a hydrogen ionization timescale that is significantly longer than its cooling timescale (e.g., \citealt{Wolfire_2003}).} We assume an optically-thin, spatially-uniform photoionising background radiation field based on the redshift $z=0$ tabulation from \cite{Haardt_2012} when computing the ionization state and cooling rates for both primordial species and metals. For gas at temperatures $T < 2 \times 10^4$ K, we also assume a photoelectric (volumetric) heating rate
\begin{align}
\Gamma_{\text{pe}} = 8.5 \times 10^{-26} \, n_{\rm H} \, \text{ergs cm}^{-3} \, \text{s}^{-1} \, ,
\end{align}
where $n_{\rm H}$ is the sum of H~\textsc{i} and H~\textsc{ii} number densities, which is the default setting for photoelectric heating in \textsc{Grackle} and agrees at order of magnitude with the solar neighbourhood value of photoelectric heating calculated by \cite{Wolfire_2003} (their Eq. 19).

The \textsc{Grackle} cooling implementation results in a three-phase neutral ISM, with a warm phase (WNM), a cool phase (CNM), and an unstable phase at intermediate temperatures. We use \textsc{Grackle} rather than the default cooling implementation in \textsc{GIZMO}, as used in, e.g., the FIRE-2 simulations \citep{Hopkins_2018}, because the default \textsc{GIZMO} cooling implementation contains an error that prevents it from producing an unstable phase of the neutral ISM, as shown in 
\aref{appendix:cooling_curve}.

Using the implementation in \textsc{GIZMO} (originally based on that of \citealt{Springel_2005b}), we form stars by stochastically converting gas particles into star particles (e.g., \citealt{Katz_1996}) such that the expectation value of the instantanous star formation rate density satisfies
\begin{align}
\langle\text{SFR}\rangle = \, \frac{d(-p \rho)}{dt} \, = \, \begin{cases}
    0 & \rho < \rho_{\text{crit}} \, , \\
    \rho \epsilon_{\star} / t_{\text{ff}} & \rho \ge \rho_{\text{crit}} \, ,
\end{cases}
\label{eq:sfr}
\end{align}
where $p$ is the probability of star formation, $\rho$ is the gas density, $\epsilon_{\star}$ is the star formation efficiency parameter, and
\begin{align}
t_{\text{ff}} = \sqrt{3 \pi / 32 G \rho} \, ,
\end{align}
is the gas free-fall timescale.
We choose the critical gas density $\rho_{\text{crit}} = 100 \, \text{H cm}^{-3}$, where this density is approximately the Jeans density for $50$ K gas at our gas particle mass resolution, and we set the star formation efficiency parameter $\epsilon_\star = 0.01$ to match the observed star formation efficiency per free-fall time in dense molecular clouds (see, e.g., Figure 10 of \citealt{Krumholz_2019} and references therein).
For densities $\ge 100$ times that of the critical density $\rho_{\text{crit}}$, we increase the star formation efficiency parameter to unity in order to avoid runaway collapse in Jeans-mass-unresolved dense regions with infinitesimal timesteps.\footnote{When using MFM/MFV methods, and \emph{unlike} in SPH, this problem cannot be solved by setting a nonzero gas particle kernel smoothing length, since the effective volume of the gas particles in these methods is determined by the nearest-neighbor distance, not the smoothing length (see \citealt{Hopkins_2015}). Setting a nonzero minimum gas smoothing length that is greater than the nearest-neighbor distance between particles when using MFM/MFV methods causes an explosive (and catastrophic) numerical instability.} We note that this star formation criterion implies that over a timestep, the star formation probability $p$ for a given gas particle takes the form 
\begin{align}
p = 1 - \exp (-\epsilon_{\star} \Delta t / t_{\text{ff}}) \, ,
\end{align}
since we must integrate \autoref{eq:sfr}, which gives a probability \emph{rate}, over a timestep $\Delta t$ in order
to obtain a probability $p$ \citep{Katz_1996}.

We implement photoionisation feedback following the method of \citet{Armillotta_2019}, which re-implements the Stromgren volume method described by \cite{Hopkins_2018}. However, because our resolution is lower than that of \citet{Armillotta_2019}, for this paper we do not use stochastic sampling from the IMF for each star particle. Instead, for simplicity, we assume that the stellar initial mass function is fully sampled, and we adopt a constant ionising luminosity of $10^{39}$ ionising photons per second per 100 \Msun \, of stellar mass from birth to 5 Myr after formation, and zero thereafter. We note that in our implementation, while a gas particle is identified as being within an H~\textsc{ii} region, the cooling and heating source terms are turned off for that particle.

For supernova feedback, we use the method of \cite{Hopkins_2018b} as implemented in \textsc{GIZMO}. In simplified form, this method couples the momentum from the unresolved Sedov-Taylor phase to the faces of the neighbouring fluid elements. Then, after subtracting the resulting kinetic energy from a fiducial explosion energy of $10^{51}$ ergs, the remaining energy is coupled to the neighbouring gas particles as a thermal energy source term, with a minimum thermal energy injection of one-half of the initial explosion energy. This method is very similar to the algorithm of \cite{Kimm_2014}, except that the GIZMO algorithm also adds the momentum of the pre-shock ejecta and has minor differences in the treatment of the difference between the simulation frame and the explosion frame. The fiducial normalisation of the maximum injected momentum from the unresolved Sedov-Taylor phase used by \textsc{GIZMO} is \footnote{The normalisation as implemented in the public source code of \textsc{GIZMO} is greater than the published normalisation value \citep{Hopkins_2018b} by a factor of $\sqrt{2}$. Additionally, \textsc{GIZMO} prior to October 2020 did not correctly normalize the momentum injected according to Eq. 18 of \cite{Hopkins_2018b}, which meant that the total momentum injected was dependent on the particle configuration surrounding a given star particle. In $\sim 1$ per cent of cases, this may lead to an unphysically-large injection of momentum.}
\begin{align}
p_{\text{SN},t} &= 6.79 \times 10^5 \, \left( \frac{E_{\text{SN}}}{10^{51} \, \text{ergs}} \right) \, f_n^{-0.14} \, f_Z^{-0.14} \, \text{\Msun} \,  \text{km} \, \text{s}^{-1} \, ,
\label{eq:terminal_momentum}
\end{align} 
where
\begin{align}
f_n(n) &= \begin{cases} 
    0.001 & n < 0.001 \\
    \frac{n}{1 \, \text{H cm}^{-3}} & n \geq 0.001
 \end{cases}
\end{align}
and
\begin{align}
f_Z &= \begin{cases} 
    0.001 & Z/Z_{\odot} < 0.01 \\
    \left( \frac{Z}{Z_{\odot}} \right)^{1.5} & 0.01 \leq Z/Z_{\odot} < 1 \\
    \frac{Z}{Z_{\odot}} & Z/Z_{\odot} \geq 1
 \end{cases}
\end{align}
This normalisation is somewhat larger than the commonly-used normalisation of $3 \times 10^5$ \Msun \, km s$^{-1}$ \citep{Thornton_1998}, but is fully consistent with the increase in momentum expected due to stellar clustering \citep{Gentry_2017, Gentry_2019, Gentry_2020}. Following the default used in \textsc{GIZMO}, we assume a constant supernova rate of $3 \times 10^{-4}$ SNe \Msun$^{-1}$ Myr$^{-1}$ for star particles with ages $0 < t_{\text{age}} < 30$ Myr. Each event is assumed to have an explosion energy of $10^{51}$ ergs. We neglect type Ia supernova explosions in our model.

\section{Simulations}
\label{section:simulations}
\subsection{Initial conditions}

We carry out two simulations: one with hydrodynamics only, and one with MHD. The initial conditions of the gas and collisionless components in both simulations are identical to those used by the AGORA isolated disc galaxy comparison
project in their `high-resolution' case \citep{Kim_2016}. These include a dark matter halo component of mass
$M_{200} = 1.07 \times 10^{12}$ \Msun \, (defined as the mass enclosed within a mean density of 200 times the critical density)
with a concentration parameter $c = 10$, a stellar disc of mass $3.4 \times 10^{10}$ \Msun,
a bulge of mass $4.3 \times 10^{9}$ \Msun, and a gas disc of mass $8.6 \times 10^{9}$ \Msun.
This implies a gas fraction of $\sim 20$ per cent by mass in the initial conditions.
The stellar disc has a scale height of approximately $350$ pc and thus represents the observed stellar `thin disc' in the Galaxy. The gas disc has a scale height of $R_0 = 3.43218$ kpc
and scale length $z_0 = 0.343218$ kpc, with the gas disc initialized with the density profile
\begin{align}
\rho(r,z) = \rho_0 \, \exp (-R/R_0) \, \exp (-|z|/z_0) \, .
\end{align}
The gas temperature and circular velocity are initially computed via solution of the Jeans equations
to be in hydrostatic and centrifugal equilibrium using the method described by \cite{Springel_2005}. However, we
override the hydrostatic temperature values with a uniform gas temperature of $10^4$ K within the disc in order to allow the disc
to rapidly cool and collapse to form stars.
After stars form, the disc is then partially re-inflated by the injection of momentum and energy from supernovae. Our simulations use gas particles with mass 859.3 \Msun, dark matter particles with mass $1.254 \times 10^5$ \Msun, and stellar disc and bulge particles with mass $3.4373 \times 10^3$ \Msun. (Star particles formed during the simulation have the mass of the gas particle from which they were stochastically converted.)

In the MHD simulation, the initial magnetic field is:
\begin{align}
B_{R} &= 0 \, ,\\
B_{\phi}(R, \phi, z) &= B_0 \, \exp(-R/R_0) \, \exp(-|z|/z_0) \, ,\\
B_{z} &= 0 \, ,
\label{eq:bfield}
\end{align}
where $B_0 = 10$ \microgauss.
This field is analytically divergence-free. 
However, when discretised onto the gas particles in the initial conditions it is not, but the residual 
numerical divergence is rapidly transported outside the domain via divergence-cleaning and local approximate Hodge projection \citep{Hopkins_2016}.
This field geometry (purely toroidal) and strength are chosen to be in rough approximate agreement with the observed `ordered' (or `regular') magnetic field of the Galaxy,
with the expectation that the turbulent component of the field (as well as any non-toroidal component) would be generated by the gravitational collapse and stellar feedback in the simulation.

\subsection{MHD simulation}

\begin{figure*}
	\includegraphics[width=\textwidth]{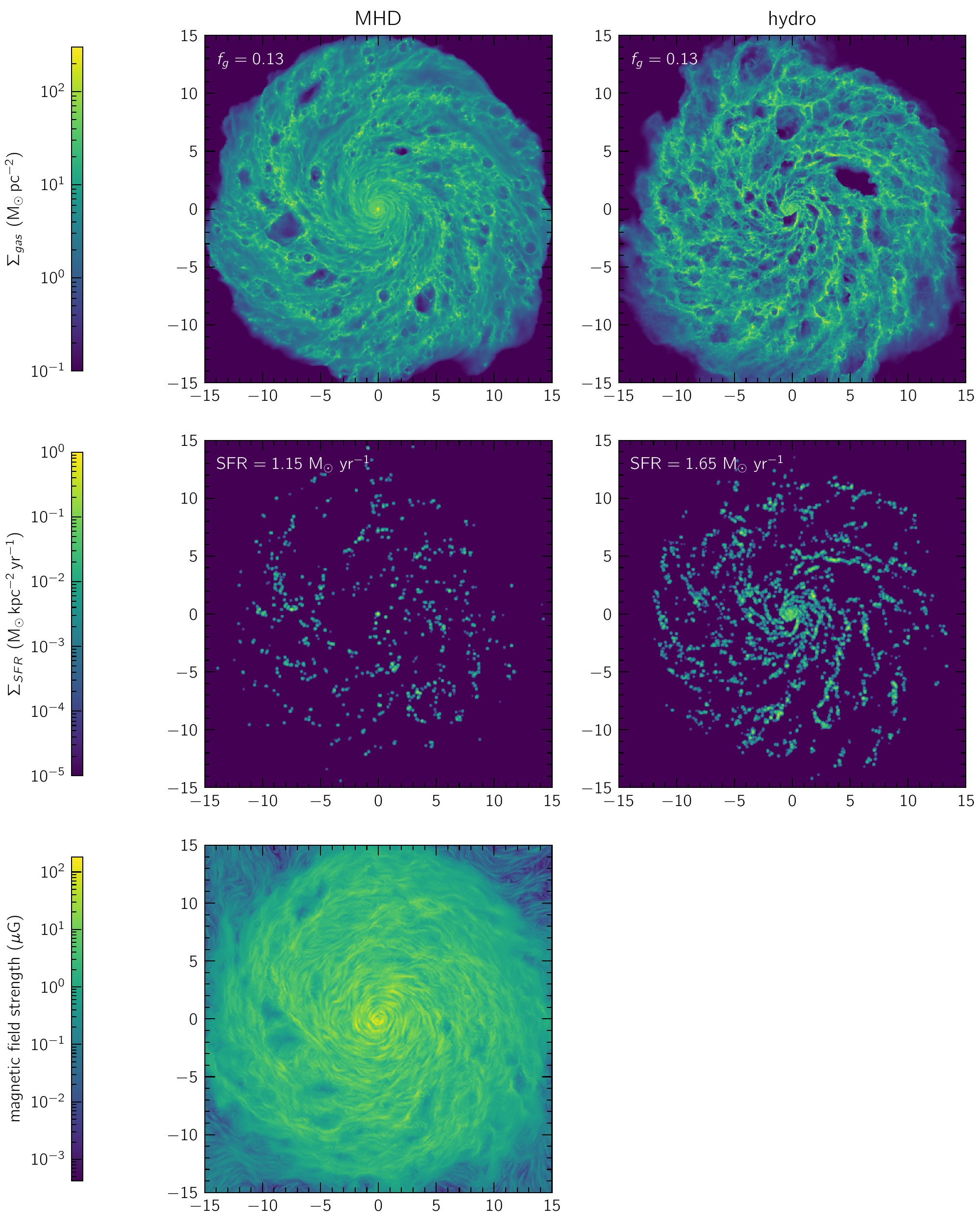}
    \caption{\emph{First row:} The face-on density projection of the MHD simulation (left) and the hydro simulation (right) at time $t \sim 1$ Gyr. \emph{Second row:} The face-on projected star formation rate surface density expectation value of the MHD simulation (left) and the hydro simulation (right). \emph{Third row:} The face-on in-plane magnetic field strength of the MHD simulation, with the field direction indicated via a line integral convolution \citep{Cabral_1993}.}
    \label{fig:faceon_density}
\end{figure*}

We run the MHD simulation for a time $t=976$ Myr.\footnote{This is a result of the fact that our dimensionless time units are derived from length units of $1$ kpc, velocity units of $1$ km s$^{-1}$, and mass units of $10^{10}$ \Msun.} The face-on projected density, the face-on projected star formation rate, and the face-on, in-plane magnetic field of the final output of the simulation are shown in the left column of \autoref{fig:faceon_density}. We see morphology typical of an S0 galaxy, with relatively weak spiral arms and no visible bar. The lack of a bar or a grand design spiral pattern may be due to a lack of close-in orbiting satellites, such as the Large Magellanic Cloud, which are absent in our simulation. The surface density normalisation is roughly consistent with observations of the Milky Way, with values in the range $50-100$ \Msun pc$^{-2}$.

The magnetic field strength generally ranges from $10-100$ \microgauss, which is also consistent with observations when compared with the \emph{total} magnetic field strength, not just the so-called `ordered' field. We have used a line integral convolution of the in-plane (vector) magnetic field applied on top of the magnetic field strength in order to illustrate the sense of the magnetic field direction (although the visualization is degenerate up to a global reversal of the direction). There do not appear to be any magnetic field reversals in the azimuthal direction (i.e., along a circular orbit), which is in tension with the common interpretation of the Galactic Faraday rotation measurements that suggest a reversal of the azimuthal magnetic field direction between spiral arms \citep{Beck_2015}. However, since our initial conditions do not have any magnetic field reversals (the initial field is everywhere aligned in the $+\phi$ direction), nor any gas accretion from halo gas or cosmological sources, the lack of field reversals may not be unexpected. The mechanism for generating field reversals is unknown, although they are seen in some cosmological simulations of magnetic fields \citep{Pakmor_2018}, and are suggested to form as a result of a mean-field dynamo process \citep{Beck_2015}.

\begin{figure*}
	\includegraphics[width=\columnwidth]{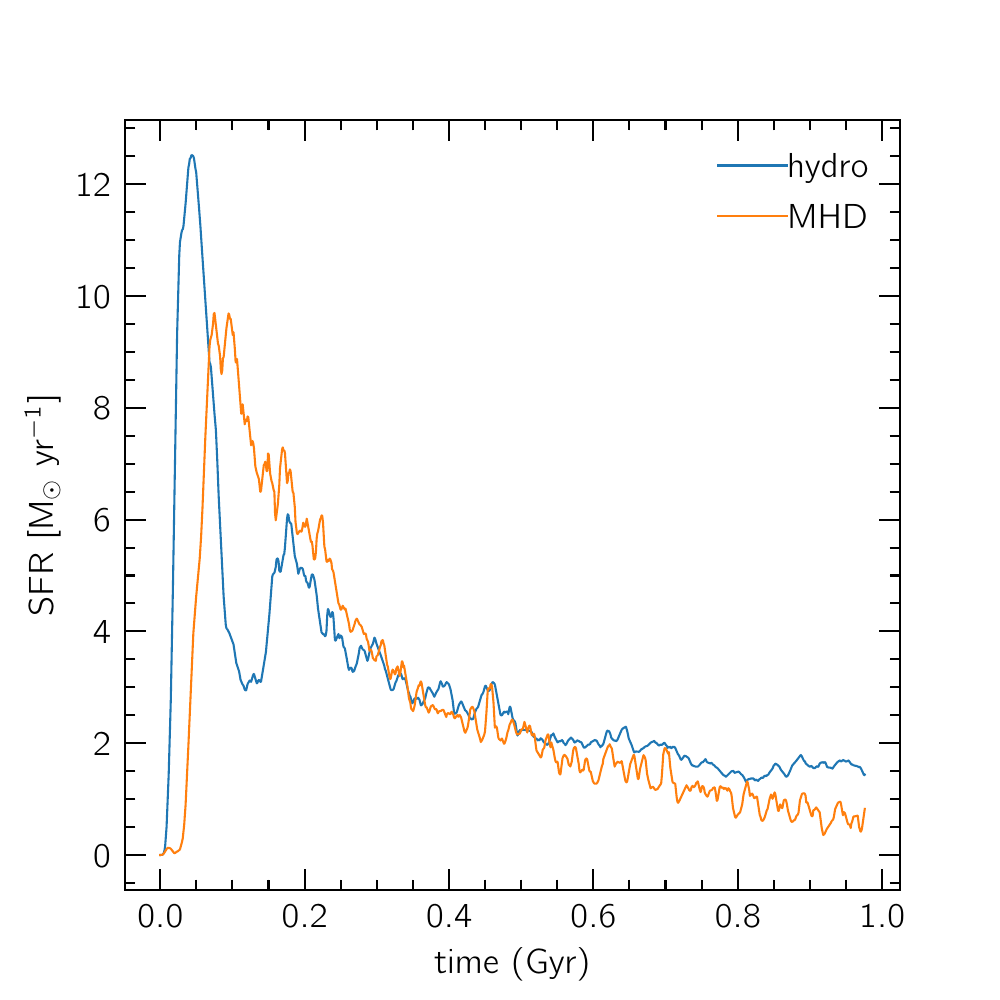}
    \includegraphics[width=\columnwidth]{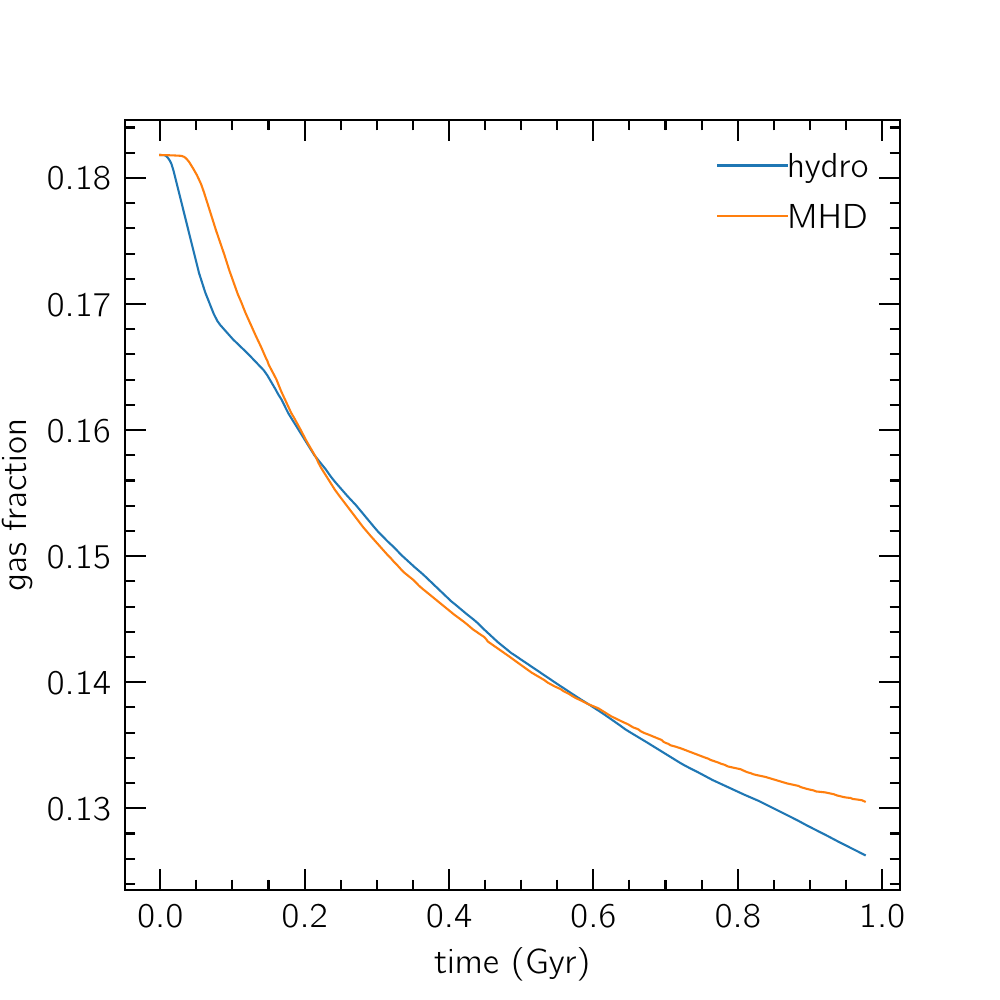}
    \includegraphics[width=\columnwidth]{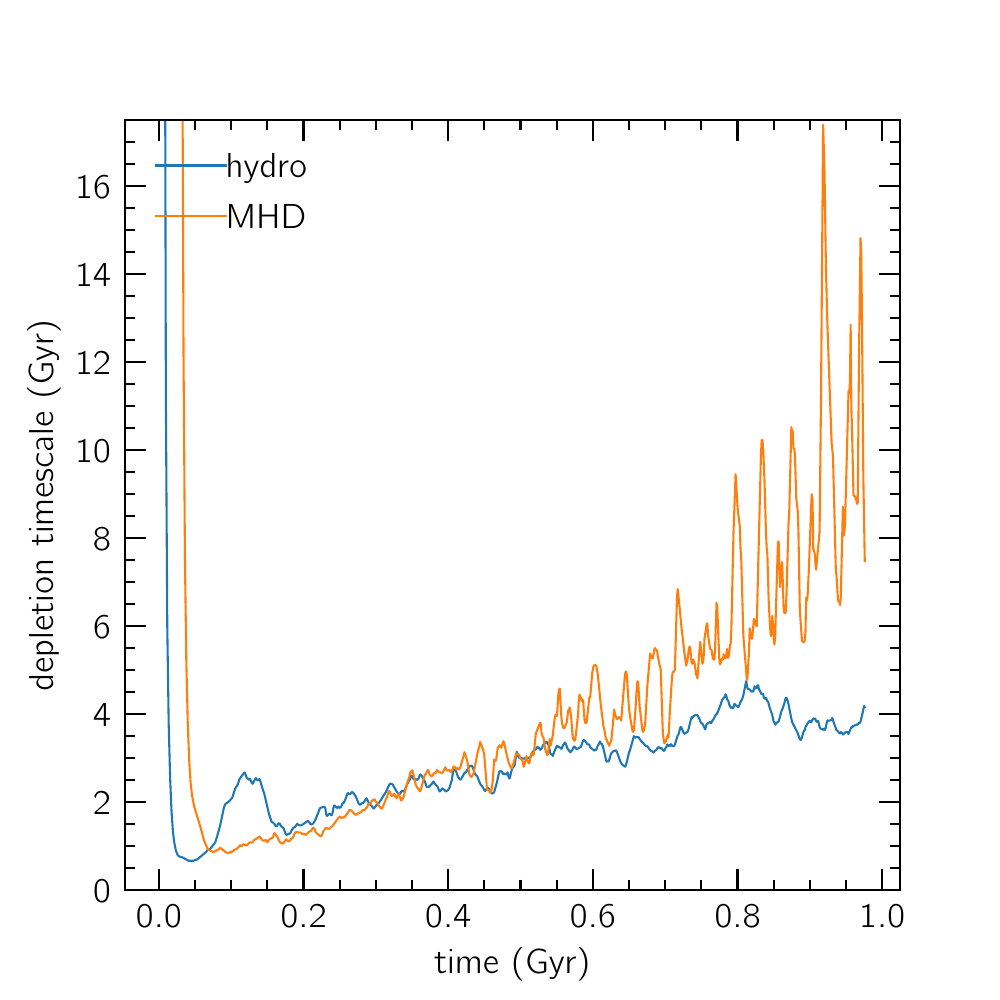}
    \includegraphics[width=\columnwidth]{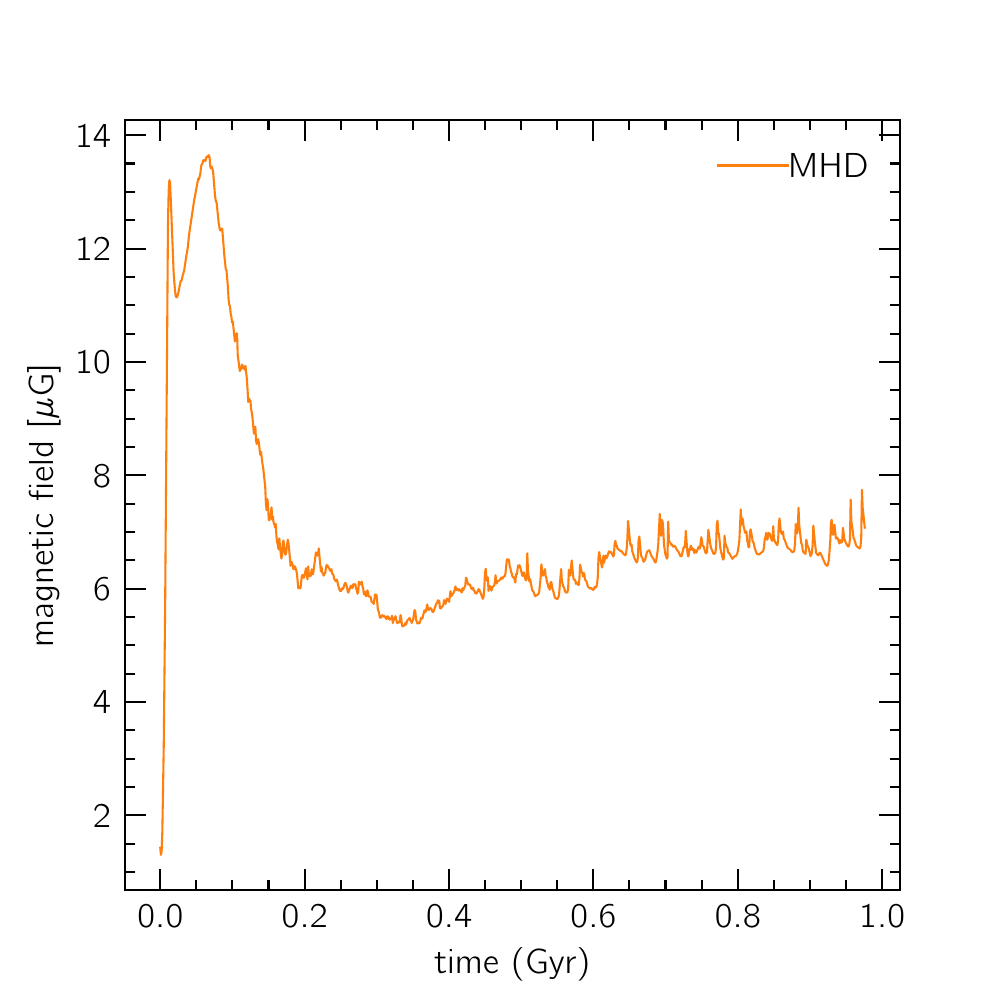}
    \caption{\emph{Top left:} The star formation rate of the simulations as a function of time. \emph{Top right:} The gas fraction of the simulations as a function of time. \emph{Bottom left:} The gas depletion timescale of the simulations as a function of time. \emph{Bottom right:} The mass-weighted mean magnetic field of the MHD simulation as a function of time.}
    \label{fig:timeseries}
\end{figure*}

In the top left panel of \autoref{fig:timeseries}, we show the star formation rate as a function of elapsed simulated time. We see that it initially spikes at a value of around $10$ \Msun \, yr$^{-1}$, as a the disc cools and there is insignificant turbulence (or any other feedback processes) to slow the collapse of gas and subsequent formation of stars. The star formation rate then slowly declines as collapse and feedback processes come into a quasi-equilibrium toward the end of the simulation, plateauing at a rate of $\sim 1-2$ \Msun \, yr$^{-1}$. As star formation proceeds, the gas is necessarily consumed. In the top right panel of \autoref{fig:timeseries}, we show the gas fraction of the disc (computed as the mass of gas particles relative to the total non-dark-matter mass). From the initial gas fraction of $\sim 0.2$, we see a slow decline over $\sim 1$ Gyr to a value of $\sim 0.1$ gas fraction. The decline in star formation rate is somewhat more rapid than the decline in star formation rate, so the total depletion time (bottom left panel of \autoref{fig:timeseries}), defined as the ratio of the gas mass to the star formation rate, very gradually increases as the simulation runs, reaching $\approx 6-8$ Gyr at the final time. This is roughly consistent with the depletion times observed in nearby spiral galaxies, where molecular gas depletion times average $\approx 2$ Gyr \citep[e.g.,][]{Leroy_2013}, but molecular gas constitutes only $\approx 1/3-1/4$ of the total gas mass (with the balance as H~\textsc{i}; \citealt{Saintonge_2011}), so the total gas depletion time is a $\approx 6-8$ Gyr.

Finally, the lower right panel of \autoref{fig:timeseries} shows the gas mass-weighted mean magnetic field of the MHD simulation. The initialisation of the magnetic field according to \autoref{eq:bfield} implies a mass-weighted mean field of $\sim 1$ \microgauss. The field is subsequently strongly amplified by the initial burst of star formation described previously, and then relaxes into a quasi-steady-state value of $\sim 7$ \microgauss.

\begin{figure*}
	\includegraphics[width=\textwidth]{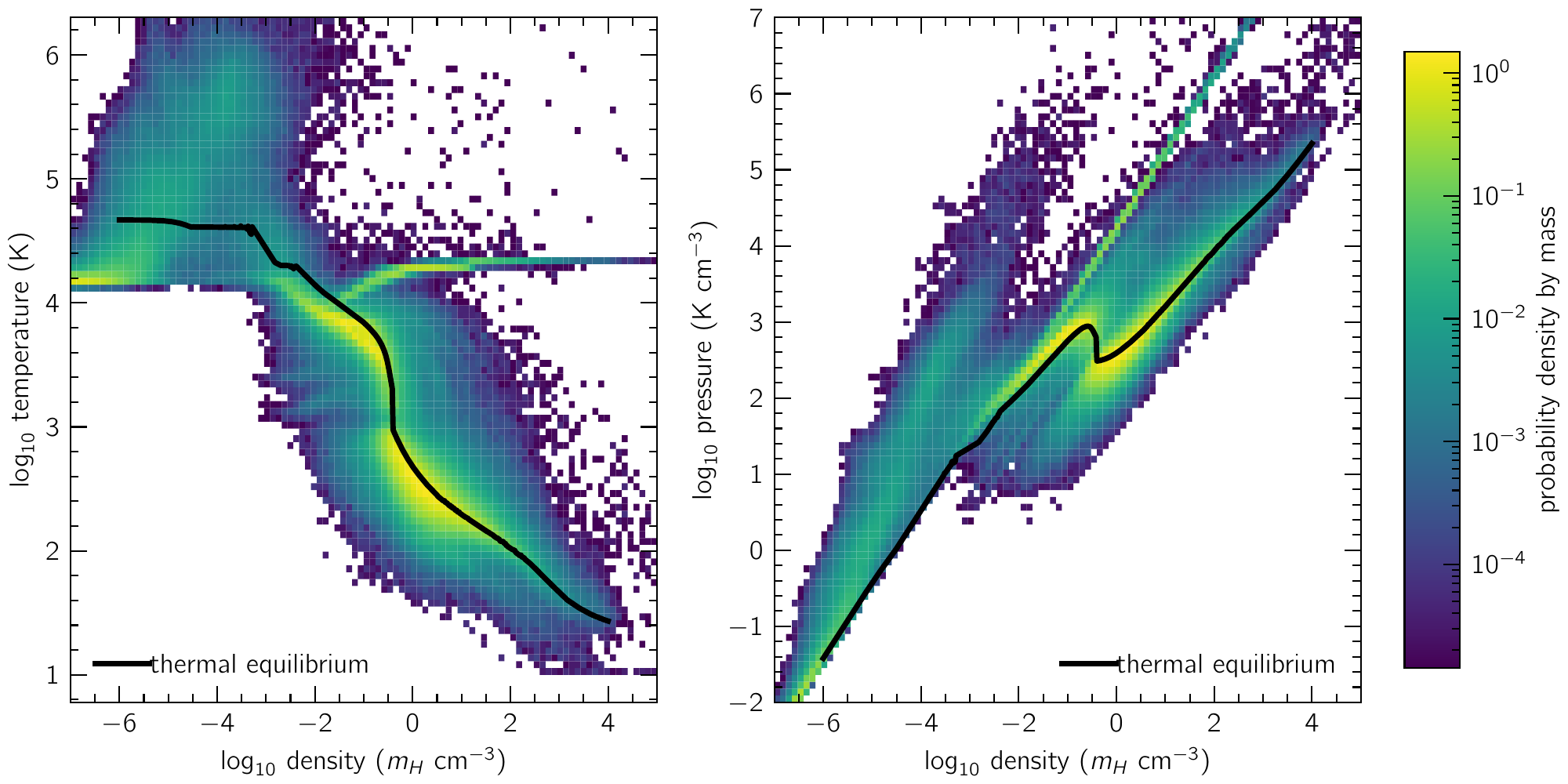}
    \caption{\emph{Left:} The temperature-density distribution of gas in the MHD simulation. \emph{Right:} The pressure-density distribution of gas in the MHD simulation.}
    \label{fig:cooling_curve_sim}
\end{figure*}

In \autoref{fig:cooling_curve_sim}, we show the mass-weighted distribution of temperature and gas density in the MHD simulation (left panel) and the mass-weighted distribution of thermal pressure and gas density in the MHD simulation (right panel). For comparison, we also overplot as solid lines the equilibrium temperature and thermal pressure as a function of density for our adopted radiative heating and cooling physics (using Grackle version 2.2; \citealt{Smith_2017}). We see that the gas in the simulation quite closely tracks the equilibrium curves, except at low densities, where the gas is far out of thermal equilibrium due to shock heating from supernova feedback, and in dense clouds that are heated to $\sim 10^4$ K due to photoionisation from stars $\lesssim 5$ Myr old, in good agreement with similar simulations of the Milky Way ISM (e.g., \citealt{Goldbaum_2016,Kim_2017}). There is also a small systematic offset from thermal equilibrium in the warm neutral medium due to the ionisation equilibrium timescale generally exceeding the cooling time in this phase, an effect anticipated from theory \citep{Wolfire_2003}. Some recent simulations (e.g., \citealt{Hopkins_2018,Gurvich_2020}) show large deviations of cold neutral gas from its thermal equilibrium state. Such behaviour is almost certainly incorrect given the extremely short cooling times of cold neutral gas (e.g. \citealt{Wolfire_2003}) and is more likely due to the error affecting the heating and cooling rates of dense gas in these simulations that we identify in \aref{appendix:cooling_curve}.

\subsection{Hydrodynamics-only simulation}
We run an additional simulation without magnetic fields as a control in order to examine the differences between the magnetic and non-magnetic simulations. For this simulation, we simply set the initial magnetic field to zero everywhere, and leave all other properties of the initial conditions as they are in the previous simulation. We use the same code settings, including using the MHD solver, in order to ensure that the differences between the two simulations are entirely due to the presence (or absence) of magnetic fields, not the numerical properties of the hydrodynamic vs. MHD solver.

Examining the bulk properties shown in \autoref{fig:timeseries}, we see that the non-magnetized simulation has a greater initial burst of star formation, peaking just above 12 \Msun yr$^{-1}$ (upper left panel). This burst is reflected in a steeper initial decline of the gas fraction (upper right panel). In contrast to the magnetized simulation, following the initial burst, the star formation rate rapidly declines, undergoes a second burst at a time $\sim 200$ Myr, and then slowly declines over the next several hundreds of Myr to a star formation rate of $\sim 2$ \Msun yr$^{-1}$. The larger initial burst but slower decline means that the gas fraction of the hydro simulation is very similar to the MHD simulation after both simulations are stopped after $t \sim 1$ Gyr (upper right panel). As a result of the slightly higher quasi-steady-state star formation rate compared to the magnetized simulation, the gas depletion timescale has a quasi-steady-state value of $\sim 4$ Gyr when the simulation is stopped, which is 1.5--2 times lower than the value for the magnetized disc.

\section{Discussion}
\label{section:discussion}
\subsection{Global magnetic structure}
\label{section:magnetic_structure}

\begin{figure*}
	\includegraphics[width=\textwidth]{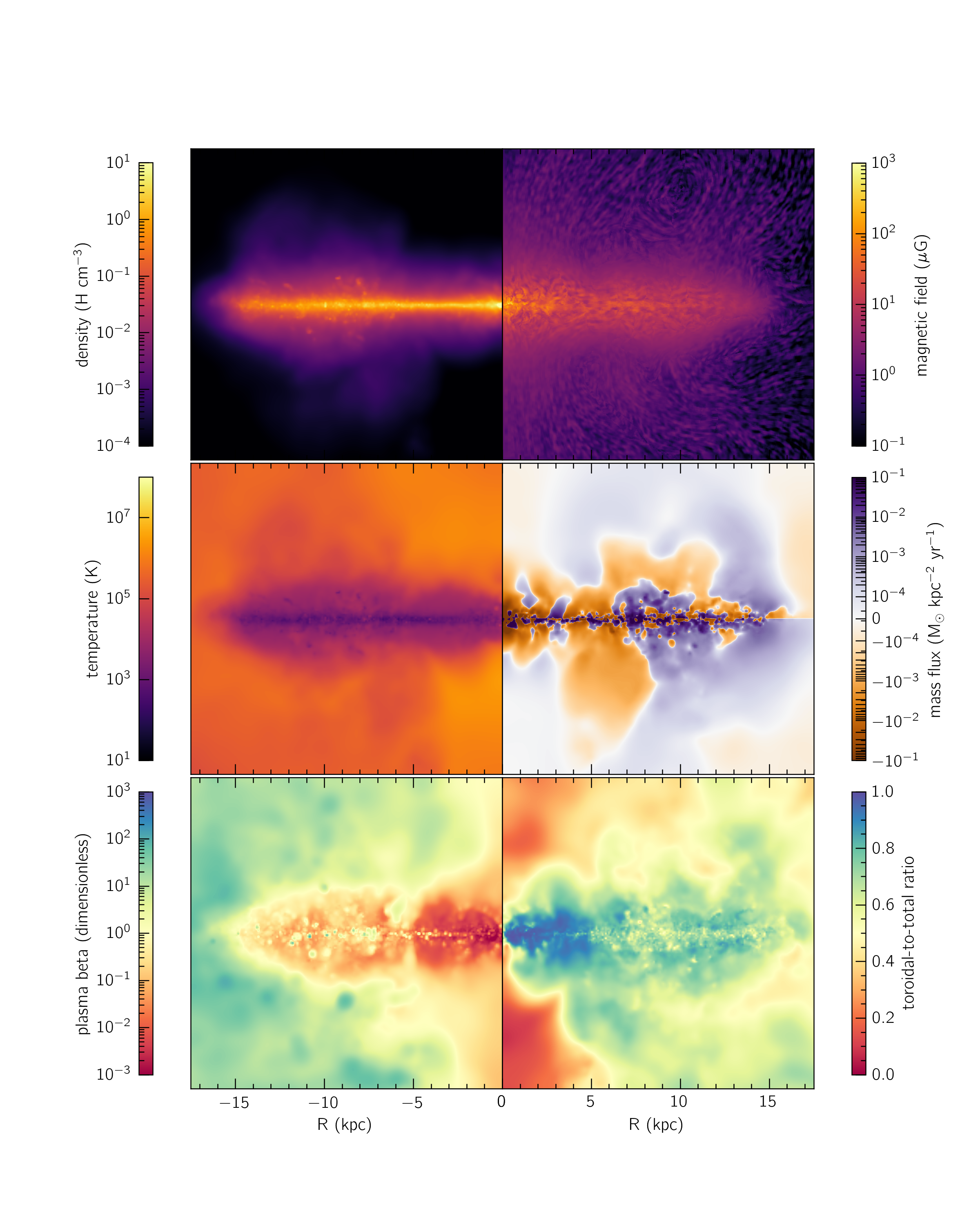}
    \caption{\emph{Upper left:} A $(R,z)$-average projection of the gas density of the magnetized simulation at simulated time $t \sim 1$ Gyr. \emph{Upper right:} A line integral convolution (LIC; \citealt{Cabral_1993}) of the $(R,z)$-average magnetic field strength convolved with the magnetic field direction in the $(R,z)$ plane. \emph{Middle left:} A $(R,z)$-average projection of the gas temperature. \emph{Middle right:} A $(R,z)$-average projection of the outgoing vertical gas mass flux. \emph{Lower left:} A $(R,z)$-average projection of the plasma beta parameter. \emph{Lower right:} a $(R,z)$-average projection of the toroidal-to-total magnetic field ratio.}
    \label{fig:rz_panels}
\end{figure*}

One of the primary goals of our study is to determine the global structure of the magnetic fields in and around Milky Way-like disc galaxies, and the relationship of the field structure to other physical quantities. To begin this investigation, in \autoref{fig:rz_panels} we show the azimuthally-averaged (i.e., in the $\phi$-direction) structure of the disc in a number of quantities; all quantities shown are mass-weighted means except for the mass flux, which is an intrinsically volumetric quantity. These projections in the radius-vertical height plane illustrate the distinct `atmospheric' components present in the disc. This separation is visible in virtually all components (except the density), including the magnetic field, the vertical mass flux, and the fraction of toroidal magnetic field. It is clear that there are three distinct zones -- a very thin disc $\sim 300$ pc in size near the midplane, a thicker, $\sim 1-2$ kpc-wide zone around that, and then a distinct third zone at larger heights. This structure is reminiscent of a multi-layered cake. Henceforth, we will refer to this stratification as the `layer-cake structure' of the disc.

In the upper left panel of \autoref{fig:rz_panels}, we see that the gas density is approximately exponentially distributed in the vertical direction, as expected. There is some substructure within the disc, as well as a few kiloparsec-scale plumes of gas above the disc as a result of supernova-driven hot outflows. The scale height of the gas density increases with galactocentric radius, indicating a `flaring' disc structure.

In the upper right panel, we see the magnetic field strength, with the direction of the field indicated via a line integral convolution \citep{Cabral_1993}. The magnetic field strength in the galactic centre is of order $100$ \microgauss, consistent with observations \citep{Beck_2015}. The field strength falls off with height and radius less steeply than the gas density, a phenomenon observed in other simulations and suggested by observations. However, unlike the gas density, the magnetic field does not vary smoothly with height. Instead, it is significantly tangled at heights of within $\sim 1$ kpc, consistent with the schematic field geometry of \cite{Boulares_1990} that was suggested by observations at the time. At large heights ($> 3$ kpc), the field becomes coherent on kiloparsec scales, with the projected field lines stretching nearly vertically out of the galaxy near zero galactocentric radius, whereas the projected field lines further away in radius ($> 5$ kpc) curve toward the disc and may form toroidal flux tubes at heights of several kiloparsecs above the disc, as seen at $R \approx 10$ kpc and $z \approx 7$ kpc in the image.

The middle left panel shows the temperature. We see a cold gas disc ($50-100$ K) in the inner $z < 300$ pc region. The cold disc is surrounded by a region of significant spatial extent, ranging from $\sim 300-500$ pc to $\sim 3$ kpc, with gas at temperatures between several thousand Kelvins and $\sim 10^4$ K. The outer part of this region can be identified with the so-called Reynolds layer \citep{Reynolds_1989} of diffuse ionised gas surrounding the Galaxy. Further out in height ($> 3$ kpc), the gas temperature is typically around $10^6$ K, originating from the hot outflows driven by supernovae, intermixed with `plumes' of rapidly-cooling intermediate temperature ($\sim 10^5$ K) gas.

The middle right panel of \autoref{fig:rz_panels} shows the vertical mass flux, which we define as $\dot{M}_z \equiv \rho v_z\mbox{sgn}(z)$, i.e., positive values indicate mass flow away from the midplane, while negative values indicate flow towards it. We use a diverging logarithmic colour scale, so that colour indicates flow direction -- outflows are purple, inflows are orange. The strength of outflows/inflows is largest at the mid-plane ($z=0$) of the galaxy, with a sharp drop-off in magnitude with height, and the direction of inflow or outflow shows many local reversals. This is suggestive of a fountain-type of outflow. At large heights ($> 3$ kpc), by contrast, we see mostly coherent outflow, with only small patches of inflow, suggesting that this region is not primarily a fountain, but instead represents a true wind of mass leaving the galaxy.

The dimensionless plasma beta ($P_{\text{thermal}} / P_{\text{magnetic}}$) is plotted in the lower left panel. We see the same basic `layer-cake' structure in plasma beta as a function of galactocentric height, but with a more pronounced intermediate region. The innermost cold gas disc is approximately equally balanced between the thermal pressure of the gas and the magnetic pressure, with the bubble-like substructure likely a result of individual H~\textsc{ii} regions and supernova remnants. Above this, there is a `corona' of magnetically-dominated gas extending to $\sim 3$ kpc, with stronger magnetic dominance toward the galactic centre. Above this corona is a gas-pressure-dominated region that is the product of hot outflows.

Finally, the lower right panel shows the ratio of the toroidal component ($B_\phi$) to the total magnetic field strength. Again, we see a zonally-stratified structure near the plane. First focus on the outer disc, $R \gtrsim 5$ kpc. In this radial region, near the plane at $z \lesssim 300$ pc, the mean toroidal fraction is $\approx 0.5$ (indicated by white on the plot), with a great deal of substructure corresponding to the positions of individual supernova remnants. Above this, at $z \sim 0.3 - 3$ kpc, is a zone where the field is $\approx 80\%$ toroidal, while at even larger height, $z \gtrsim 3$ kpc, the field becomes much less toroidal. The same layering is evident closer to the galactic centre, at $R \lesssim 5$ kpc, except that each of these zones is thinner, so the transitions between them happen at smaller $z$. At the very centre of the galaxy, $R \lesssim 100$ pc, the field is almost purely poloidal at all heights.

\subsection{Magnetic effects on supernova-driven winds}

\begin{figure*}
	\includegraphics[width=\columnwidth]{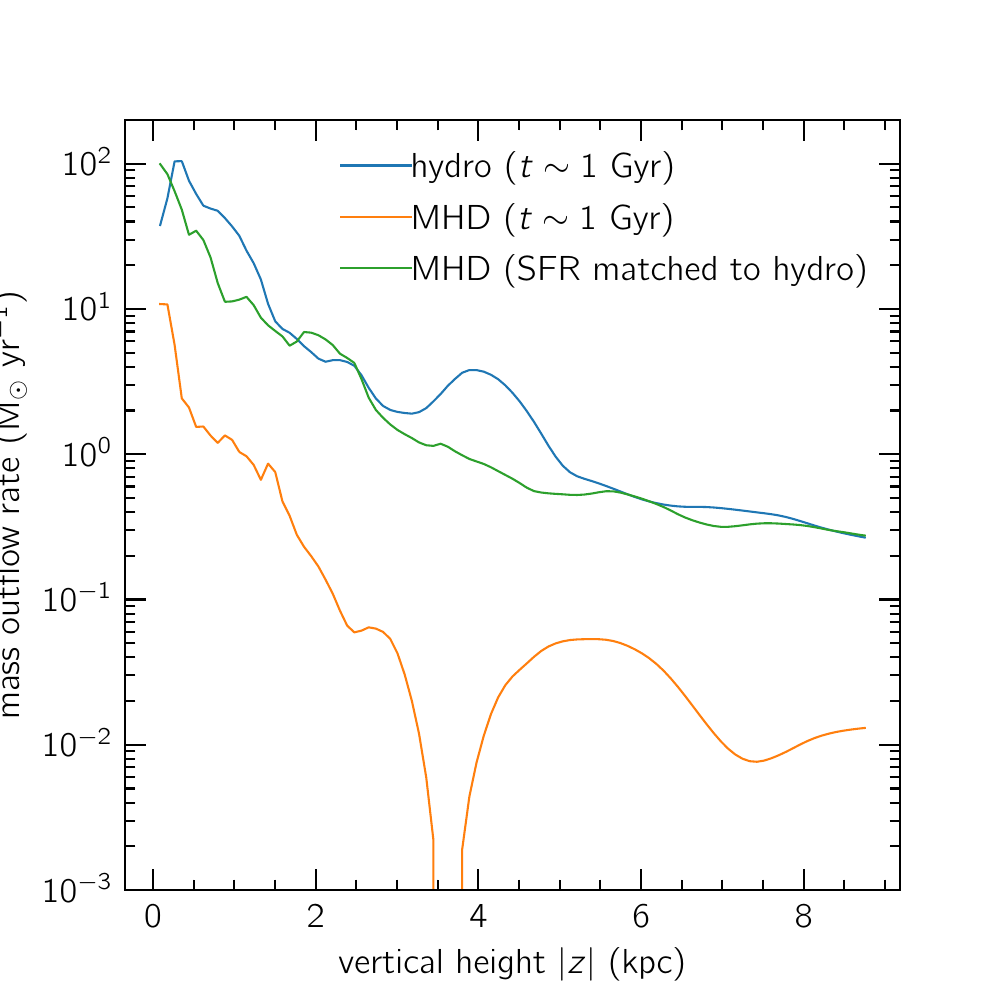}
	\includegraphics[width=\columnwidth]{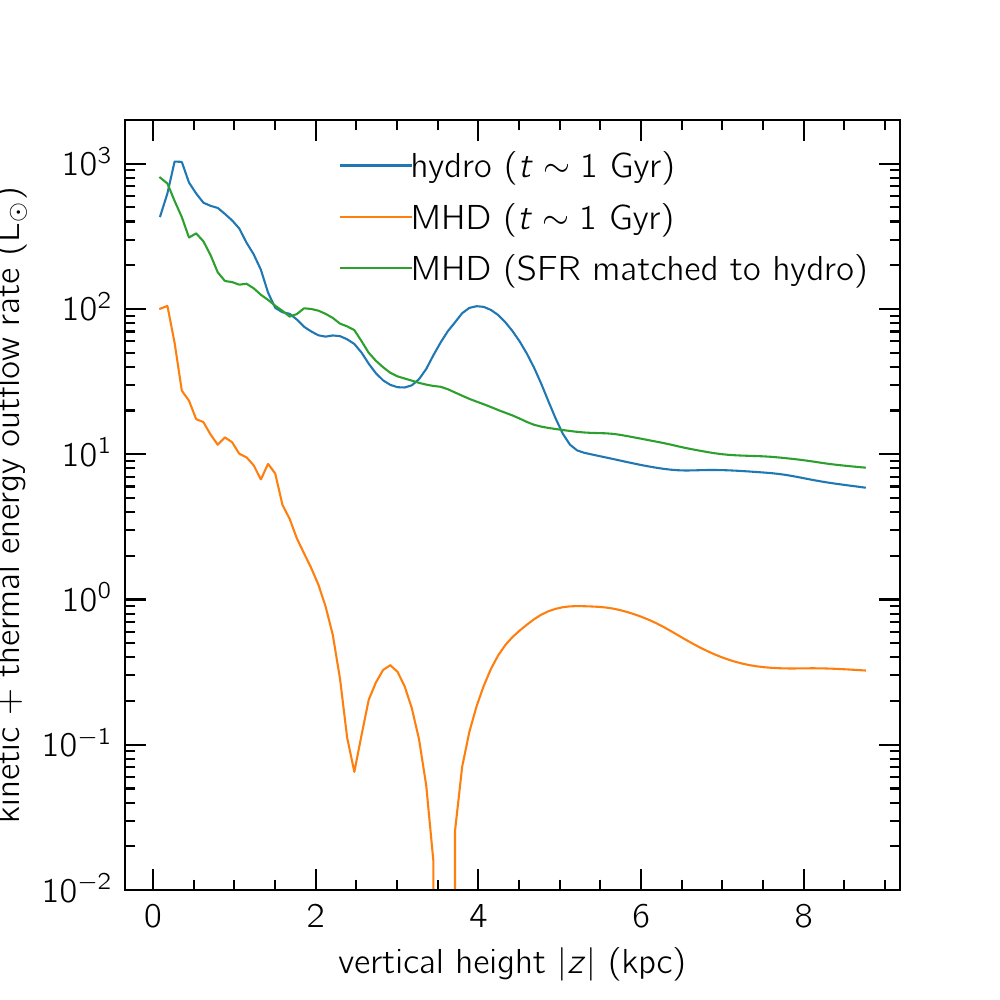}
    \caption{\emph{Left:} The vertical mass outflow rate as a function of height above the disc. \emph{Right:} The vertical kinetic and thermal energy outflow rate as a function of height above the disc.}
    \label{fig:massflux_profile}
\end{figure*}

\begin{figure*}
	\includegraphics[width=\columnwidth]{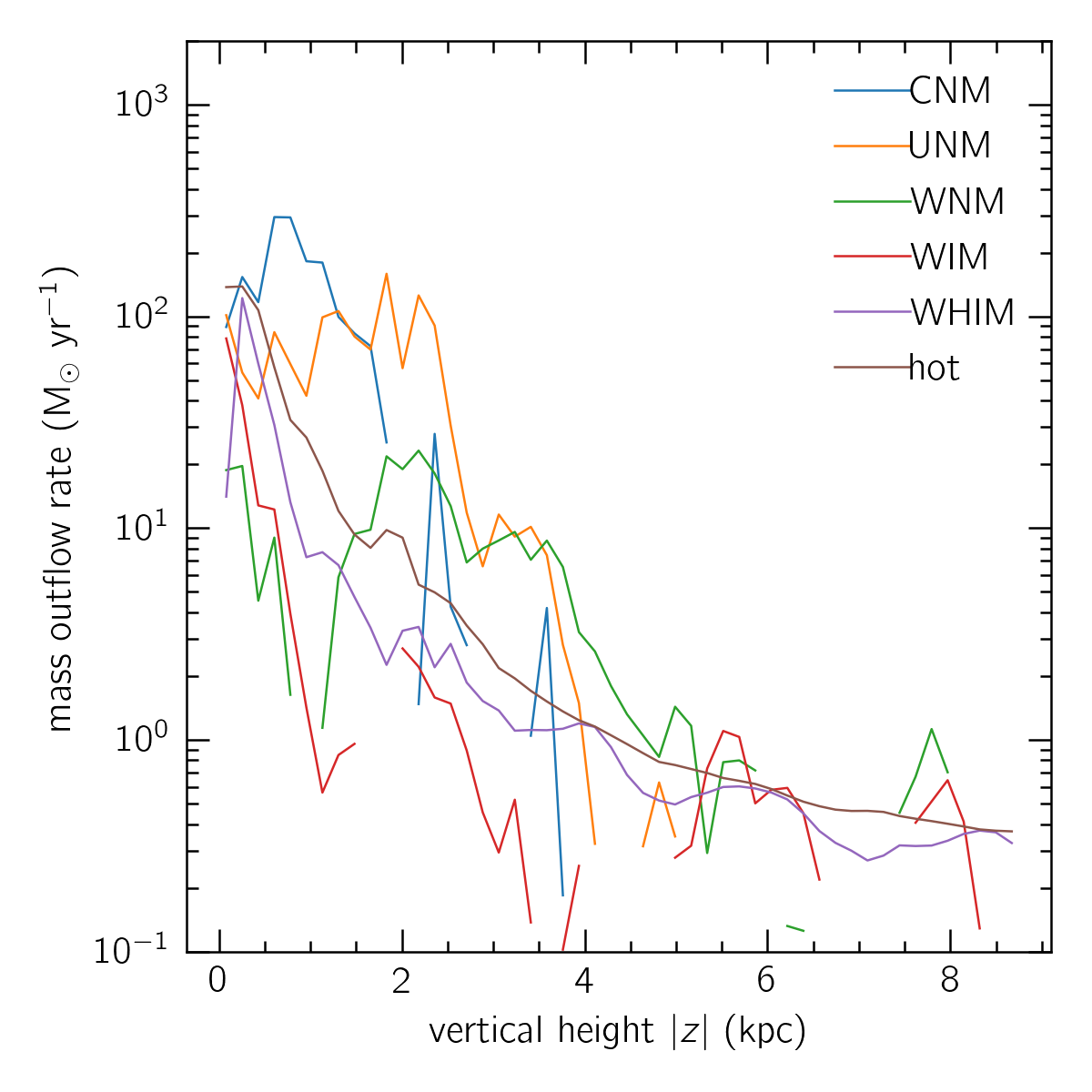}
	\includegraphics[width=\columnwidth]{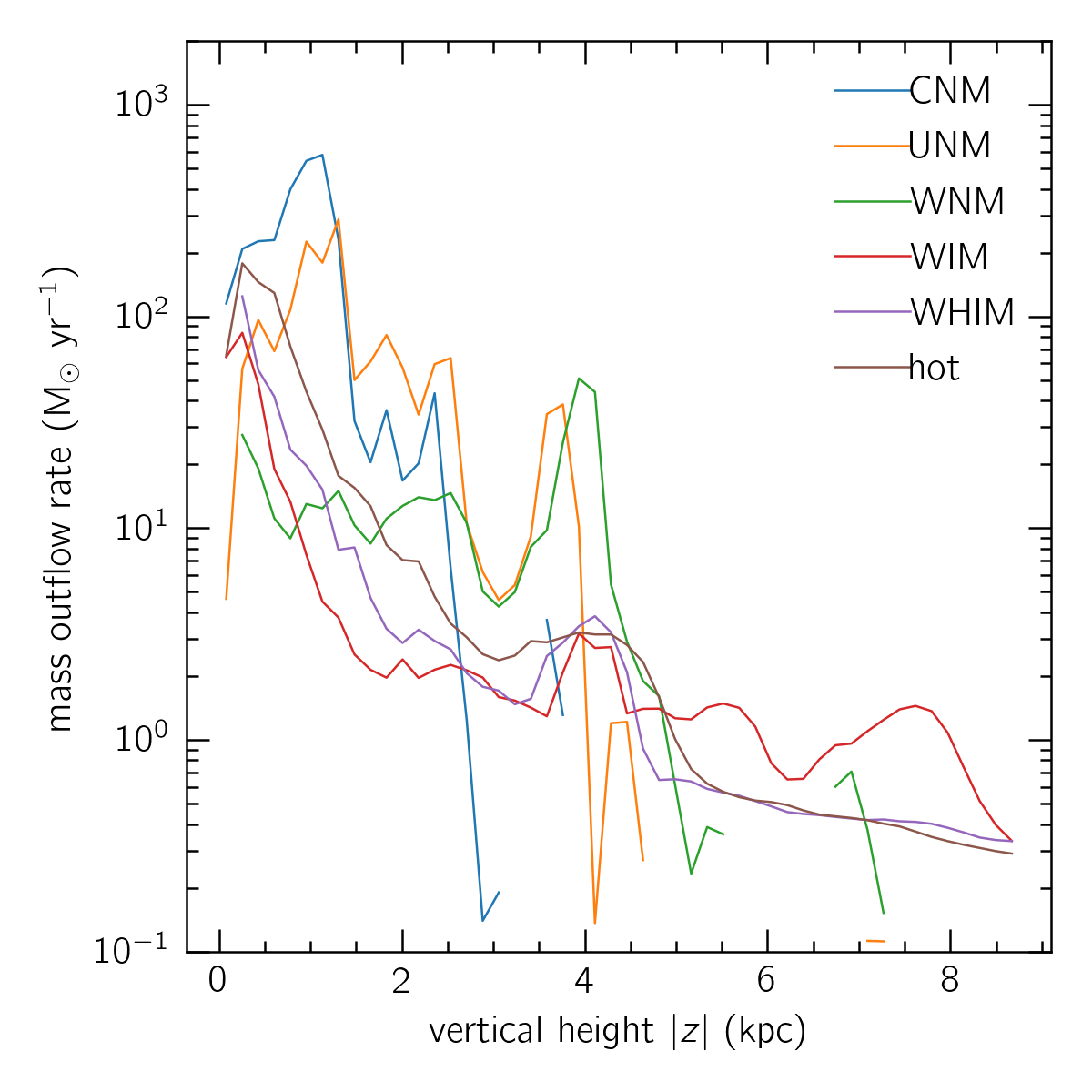}
    \caption{The vertical mass outflow rate for each thermal phase as a function of height for the magnetized (\emph{left}) and non-magnetized (\emph{right}) simulations.}
    \label{fig:massflux_phases_profile}
\end{figure*}

\begin{figure}
    \includegraphics[width=\columnwidth]{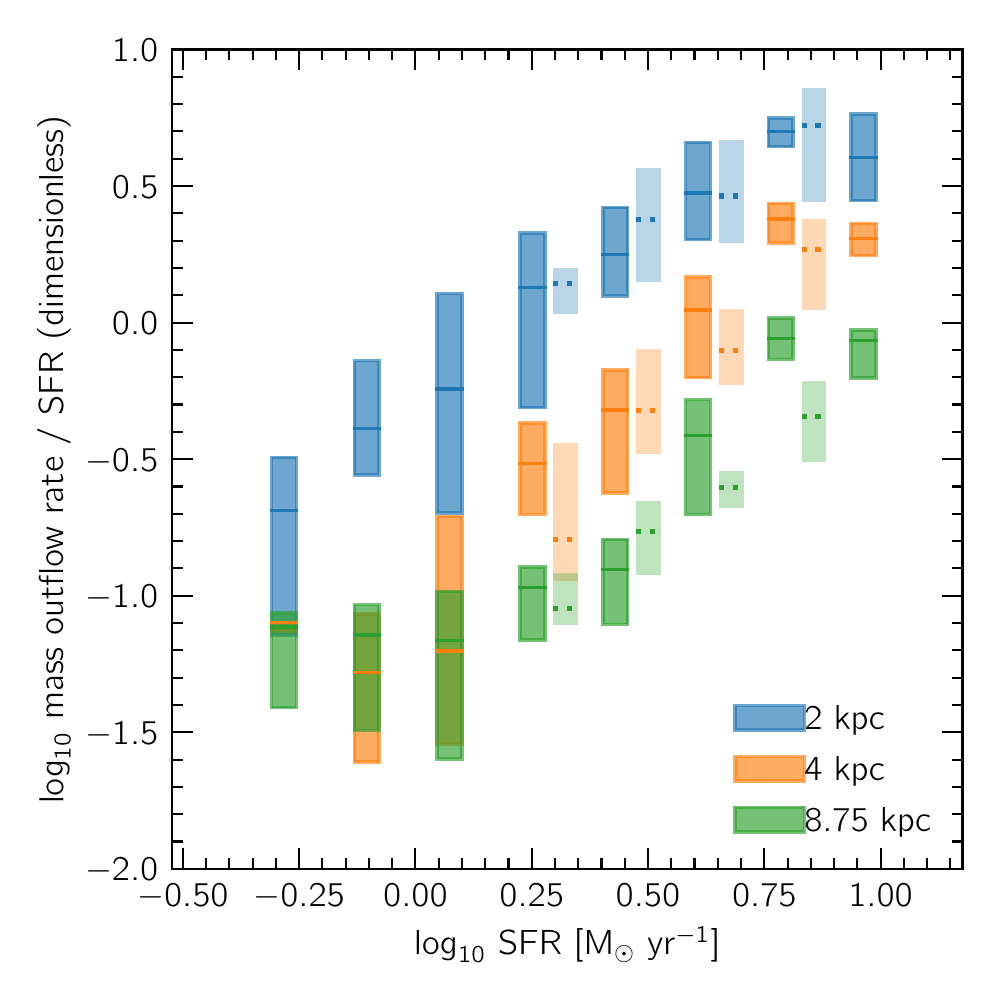}
    \caption{A box plot of the mass loading factor, computed from the instantaneous vertical mass outflow rate and the 10 Myr-averaged star formation rate, as a function of the (10 Myr-averaged) star formation rate for the MHD simulation (solid boxes and lines) and the hydro simulation (dotted lines and boxes). The boxes show the 25th, 50th, and 75th percentile of the $\log_{10}$ mass loading factor in a given bin of star formation rate. The blue boxes correspond to the outflow measured at $|z| = 2$ kpc, the orange boxes are measured at 4 kpc, and the green are measured at 8.75 kpc above/below the galactic disk.}
    \label{fig:sfr_vs_massflux}
\end{figure}

We next compare the vertical outflows of gas between the MHD and non-MHD versions of our simulations. In \autoref{fig:massflux_profile} (left panel), we show the vertical mass flux going away from the disk as a function of height, comparing the hydro simulation at $t \sim 1$ Gyr with the MHD simulation at the same simulated time and also with the MHD simulation at the same star formation rate (at an earlier simulated time). When comparing the two versions at the same simulated time, we find a substantially lower mass outflow rate at all heights in the MHD simulation (suppressed by 1-2 orders of magnitude). However, when the MHD and non-MHD simulations are matched at the same star formation rate (approximately $1.65$ \Msun yr$^{-1}$), there is no appreciable difference between the mass outflow rates as a function of height. (The large bump at $\sim 4$ kpc in the hydro outflow rate is due to a transient star formation event and has propagated outward as a traveling wave, an effect apparent in the time-dependent version of this figure.)

In \autoref{fig:massflux_profile} (right panel), we examine the vertical energy flux as a function of height above the galactic disc for both the magnetic and non-magnetic simulations. We compute the total kinetic and thermal energy in the outflow, neglecting the magnetic energy for a consistent comparison between the magnetised and non-magnetised simulations. The results are qualitatively identical to the mass outflow results, with the MHD simulation having a suppressed energy outflow rate with respect to the non-magnetised outflow rate when the two are compared at the same time, but with this effect disappearing after controlling for the star formation rate.

We further examine the mass outflow rate by decomposing it into thermal phases, including the three phases of the neutral ISM (cold neutral, unstable, and warm neutral; e.g. \citealt{Wolfire_2003}), the warm ionised medium (WIM), the warm-hot ionised medium (WHIM), and a hot phase comprised of material at all higher temperatures, as shown in \autoref{fig:massflux_phases_profile}. The boundary between cold, unstable, and warm phases is determined by finding the zero-crossings of the derivative of the equilibrium thermal pressure with respect to gas density $dP/dn$, as explained in \aref{appendix:cooling_curve}. The boundary between warm neutral and WIM is determined by finding the temperature at which the number of free electrons is one-half the number of hydrogen nucleons in thermal equilibrium. For our cooling function, we find this occurs at a temperature of 7105 K.\footnote{A somewhat higher threshold of 0.9 free electrons per hydrogen yields an ionisation temperature of $1.51 \times 10^4$ K and does not change our conclusions regarding the thermal structure of our simulations.} We set a somewhat arbitrary temperature threshold between the WIM and WHIM at $2 \times 10^4$ K, and set the boundary between the WHIM and the `hot' phase at $5 \times 10^5$ K. These choices are designed to allow the WHIM to encompass the material near the peak of the radiative cooling rate $\Lambda(T)$ of interstellar gas in collisional ionisation equilibrium at $T \sim 10^5$ K (e.g. Figure 8 of \citealt{Sutherland_1993}).  

We find that the phase structure of the outflow is not qualitatively different between the magnetised and non-magnetised simulations, and that all phases decline rapidly with height in both simulations, suggesting a fountain type of outflow. Neutral phases dominate in the fountain region $z\lesssim 3$ kpc, while ionised gas dominates at larger heights. The localised peaks and troughs apparent in the profiles are of the same order of magnitude as the temporal variability in the profiles on $\sim$ Myr timescales and we caution against over-interpreting differences between the MHD and non-MHD simulations.

In \autoref{fig:sfr_vs_massflux}, we show the (dimensionless) mass loading factor $\eta$, defined as
\begin{align}
\eta = \frac{\text{mass outflow rate}}{\text{star formation rate}} \, ,
\end{align}
as a function of star formation rate. The mass outflow rate is the instantaneous value, while the star formation rate is averaged over the previous 10 Myr. Each point represents the galaxy-averaged value a simulation output, and we compute points for each simulation snapshot (output at $\Delta t = 1$ Myr intervals) after $t = 97.6$ Myr. The color of the points indicates at what galactocentric heights $\pm z$ we computed the mass outflow. We observe a linear trend, i.e. the mass loading factor is roughly proportional to the star formation rate. This trend is the same regardless of whether magnetic fields are present in the simulation (solid boxes correspond to the MHD simulation, dotted boxes correspond to the hydro simulation). This implies that there is an approximately quadratic dependence of mass outflow rate on star formation rate, regardless of the height at which the outflow is measured. Consistent with the picture of a fountain outflow, the mean trendline for each set of measurements for various galactocentric heights shows that net mass outflow rate declines with height. This sharply disagrees with the scaling found by \cite{Muratov_2015}, who found a linear relationship between mass outflow rate and star formation rate (implying a constant mass loading factor with SFR; their Figure B1), although they measured the mass outflow rate at $0.25 R_{\text{vir}} \approx 50$ kpc (comoving) and found a mass outflow rate of $\sim 10$ \Msun yr$^{-1}$ at a star formation rate of $1$ \Msun yr$^{-1}$ at redshifts $z \sim 2-4$ for progenitors of Milky Way-mass galaxies. The significance of this disagreement is difficult to interpret, since our simulations are both non-cosmological and much higher resolution than those of \cite{Muratov_2015}, but it is possible that the scaling properties of galactic outflows may strongly differ when measured near the galaxy versus near the virial radius, or may be a strong function of redshift (or gas fraction).

The qualitative trends of our fountain outflows broadly agree with those found in the simulations of \cite{Kim_2018} and \cite{Kim_2020}. However, we find that the structure of our outflows is significantly more extended in the vertical direction, and that at $\sim 1$ kpc scales, all phases contribute a net mass outflow on average, whereas the simulations of \cite{Kim_2018} find that only the `ionized' and `hot' phases contribute to the net outflow at scales $\gtrsim 1$ kpc. At a fixed height of $\sim 2$ kpc and a comparable star formation rate, we also find a mass loading factor that is approximately an order of magnitude greater than found by \cite{Kim_2018} and \cite{Kim_2020}. This is a significant discrepancy and more work is needed in order to determine its cause. One possible explanation is that our simulations include a model for `pre-supernova' feedback in the form of photoionisation, while \citet{Kim_2018} and \citet{Kim_2020} include supernovae as their only form of spatially-localised feedback. A number of authors have found that including pre-supernovae feedback can significantly alter the properties of galactic winds, by transforming the environment into which supernova energy and momentum are deposited (e.g., \citealt{Agertz_2013}, \citealt{Kannan_2020}, Jeffreson et al.~2021, submitted). Another possible explanation is that the local box geometry of \citet{Kim_2020} does not allow streamlines to open up, which prevents the outflow from reaching the sonic point in the classical superwind solution of \cite{Chevalier_1985}, as emphasized by \cite{Martizzi_2016}.

\cite{Kim_2020} additionally find a \emph{negative} trend of mass loading factor with star formation rate, although this relationship was obtained by fitting models with significantly varying initial gas surface densities, and the same negative trend is obtained by fitting the mass loading factors and the initial gas surface densities of their models (their Figure C1). At fixed gas surface density, they likewise find a positive power-law relationship between mass loading factor and star formation rate for most of their models, with the strength of this relationship varying with initial gas surface density (their Figure 8). We leave a more detailed exploration of the outflow properties and scalings with galaxy parameters to future work.

\subsection{Implications for cosmic rays and radio observations}

Our simulated Galactic magnetic field structure has implications for the transport of cosmic rays within and out of the Galaxy. In many analytic models of cosmic rays in the Galaxy, there is a distinction between an inner region of `tangled' field lines and an outer region several kiloparsecs above the Galactic midplane with large-scale coherent magnetic fields (e.g. in the hydrostatic model of \citealt{Boulares_1990}, the wind model of \citealt{Breitschwerdt_1991}; the `base radius' of cosmic-ray-driven winds in \citealt{Quataert_2021}). Our results in \autoref{section:magnetic_structure} indicate that this transition occurs at $\sim 3$ kpc. Due to the theoretical uncertainties in the cosmic ray diffusion coefficient as used in these models, the resulting predictions for mass outflow and gamma-ray luminosity (e.g., \citealt{Lacki_2011}) cannot be used to directly test the magnetic field structure of our simulations, but our results provide a justification for a $\sim 3$ kpc value of the launching radius in a cosmic-ray-driven wind model of the Galaxy.

Our results about the vertical structure of the magnetic field may be more directly tested via spatially-resolved synchrotron emission, which is primarily sensitive to the magnetic field strength. Radio synchrotron observations typically find disk galaxies (of all inclinations) to have a scale length of $\sim 3-5$ kpc (see \citealt{Beck_2015} and references therein).

Another probe of the magnetic field are the polarized dust emission maps observed with the \emph{Planck} satellite \citep{Planck_XX}. Assuming the standard radiative torque alignment model \citep{Davis_1951,Lazarian_2007} and uniform dust properties across the Galaxy, these maps probe the plane-of-sky magnetic field orientation integrated along the line of sight. MHD simulations of local volumes of the ISM have been compared to the \emph{Planck} maps (e.g., \citealt{Planck_XX}, \citealt{Kim_2019}) but comparisons with MHD simulations with star formation, supernova feedback, and a global disk geometry are currently lacking. As a test of the realism of our simulations, however, such comparisons are somewhat limited by uncertain dust physics.

Undoubtedly the best observational comparison to our results will be the dense forest of quasar and pulsar Faraday rotation measures observable with the forthcoming Square Kilometre Array (SKA) to be completed in Western Australia \citep{Haverkorn_2015}. These measurements of the line-of-sight Galactic magnetic field will enable direct tests of MHD simulations of the Galaxy via comparison with Faraday rotation maps (Jung et al., in prep) and, in the plane of the Galaxy, tomographic mapping of the line-of-sight magnetic field.

\section{Conclusions}
\label{section:conclusions}
Our first main result is the `layer-cake' vertical structure of the magnetic field, evident in the magnetic field itself, the toroidal-to-total field ratio, the vertical gas mass flux, and most prominently, in the plasma beta as a function of Galactocentric height (\autoref{fig:rz_panels}). This structure decomposes into three approximate zones: a very thin disc $\sim 300$ pc in size near the midplane, a thicker, $\sim 1-2$ kpc-wide zone around that, and then a distinct third zone at larger heights. We emphasize the dynamical importance of this structure: in the innermost zone, the plasma beta is of order unity (magnetic and thermal gas pressure are comparable), in the intermediate zone, the plasma beta is much less than unity (magnetic pressure dominates), and in the outer zone, the plasma beta is much greater than unity (thermal gas pressure dominates). A similar zonal structure has been noted in the Galactic synchrotron emissivity, usually with a two-component structure of characteristic heights $\sim 200$ pc and $\sim 1.5$ kpc (e.g. \citealt{Boulares_1990}). Our simulation provides a theoretical picture consistent with these observations and, more importantly, suggests that the role of magnetic fields may be dynamically dominant in the intermediate zone between $300$ pc $\lesssim |z| \lesssim 3$ kpc.

Our second main result is the order unity effect of magnetic fields on the star formation rate (\autoref{fig:timeseries}). We observe that the star formation rate is suppressed by a factor of $1.5-2$ when magnetic fields of strength comparable to those observed in the Galaxy are present. However, when controlling for star formation rate, the mass outflow rate (both total and decomposed by phase) is indistinguishable between the magnetized and non-magnetized simulations (\autoref{fig:massflux_profile} and \autoref{fig:massflux_phases_profile}). The mass outflow decomposed by phase is highly stochastic and is difficult to compare with precision between the simulations. Future studies are needed in order to quantify the residual differences at better than order-of-magnitude between outflows of magnetized and non-magnetized Galactic discs.

Lastly, we obtain a positive linear correlation between mass loading factor and star formation rate in our simulations when measuring the mass outflow on kiloparsec scales above the disk (\autoref{fig:sfr_vs_massflux}). Previous work using a similar supernova feedback model found no correlation between these two quantities for gas-rich disks when measuring the mass outflow at significantly larger scales \citep{Muratov_2015}. The apparent discrepancy should be examined by future work. In searching for a theory of this enhanced mass loading, it may be productive to examine the full distribution of gas densities in which supernova explode as a function of star formation rate, since this enters as an explicit factor in our adopted feedback model (\autoref{eq:terminal_momentum}) as a result of the density dependence of the radiative cooling of supernova remnants \citep{Thornton_1998}.

\section*{Acknowledgements}
We thank L. Armillotta for providing the code to compute the HII regions in our simulations and thank N. McClure-Griffiths for useful discussions.

This research was supported by the Australian Research Council through its Discovery Projects and Future Fellowship Funding Schemes, awards DP190101258 and FT180100375. This research was undertaken with the assistance of resources and services from the National Computational Infrastructure (NCI), which is supported by the Australian Government.

The simulations in this work were run at a datacentre using 100 per cent renewable electricity sources.\footnote{For details, see \url{https://www.environment.act.gov.au/energy/cleaner-energy/renewable-energy-target-legislation-reporting}.}

\emph{Software:} matplotlib \citep{Hunter:2007},
scipy \citep{2020SciPy-NMeth},
pandas \citep{reback2020pandas,mckinney-proc-scipy-2010},
numpy \citep{harris2020array},
h5py \citep{h5py_package},
Snakemake \citep{snakemake:2021},
lic \citep{lic_package},
Meshoid \citep{meshoid_package},
GIZMO \citep{Hopkins_2015},
Grackle \citep{Smith_2017}.

\section*{Data Availability}

Due to the large volume of data products produced (approximately $4.6$ terabytes), a limited subset of the raw simulation outputs (and their associated processed data products) are permanently archived as an open-access Zenodo dataset \citep{wibking_benjamin_d_2021_4744917}. Two simulation snapshots at different simulated times ($t \sim 556$ Myr and $t \sim 976$ Myr) are included from the magnetohydrodynamic (MHD) simulation and one simulation snapshot from the hydrodynamic simulation ($t \sim 976$ Myr) is included in Gadget-2/GIZMO HDF5 format, along with processed data products from each of the simulation snapshots in NPZ (NumPy array archive) format and visualizations of the processed data products as images in PNG format.  Contingent on the ongoing availability of storage resources, additional simulation snapshots are available upon reasonable request to the authors.



\bibliographystyle{mnras}
\bibliography{bfields} 



\appendix
\section{Cooling curve comparison}
\label{appendix:cooling_curve}

\begin{figure*}
	\includegraphics[width=\columnwidth]{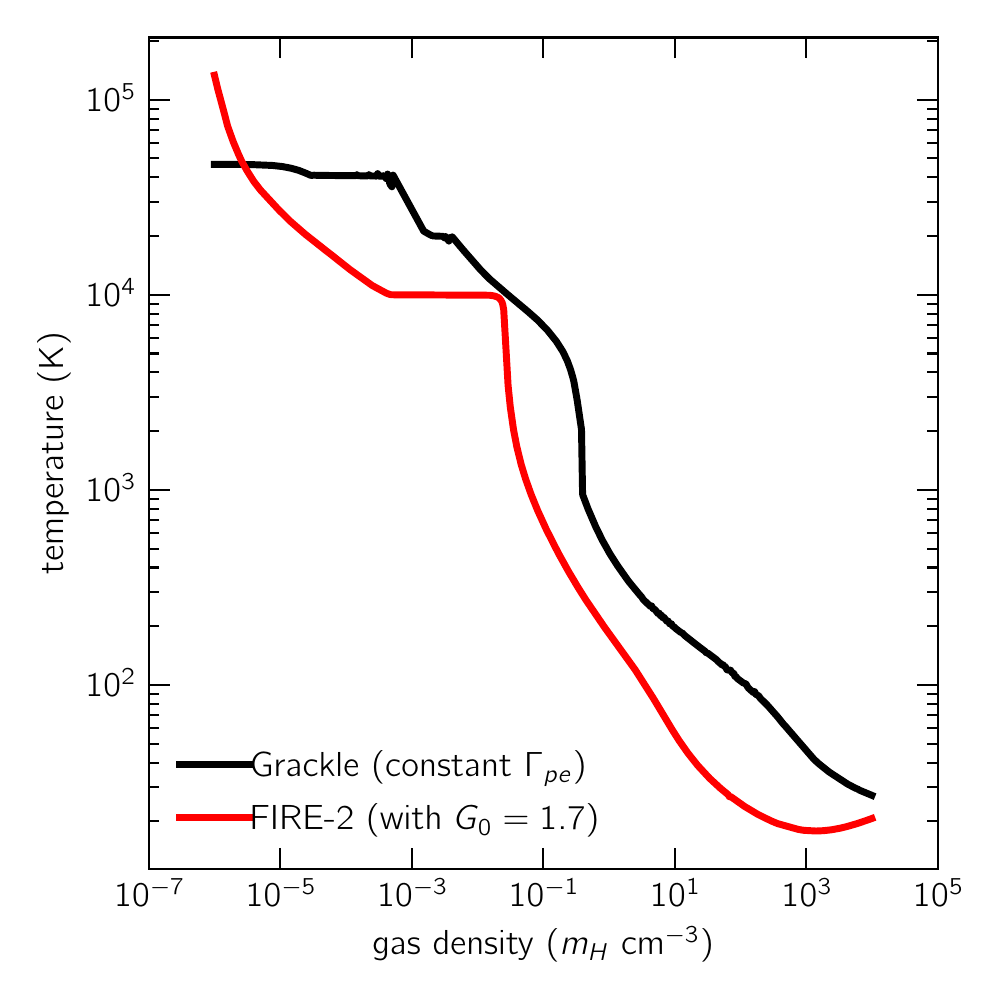}
	\includegraphics[width=\columnwidth]{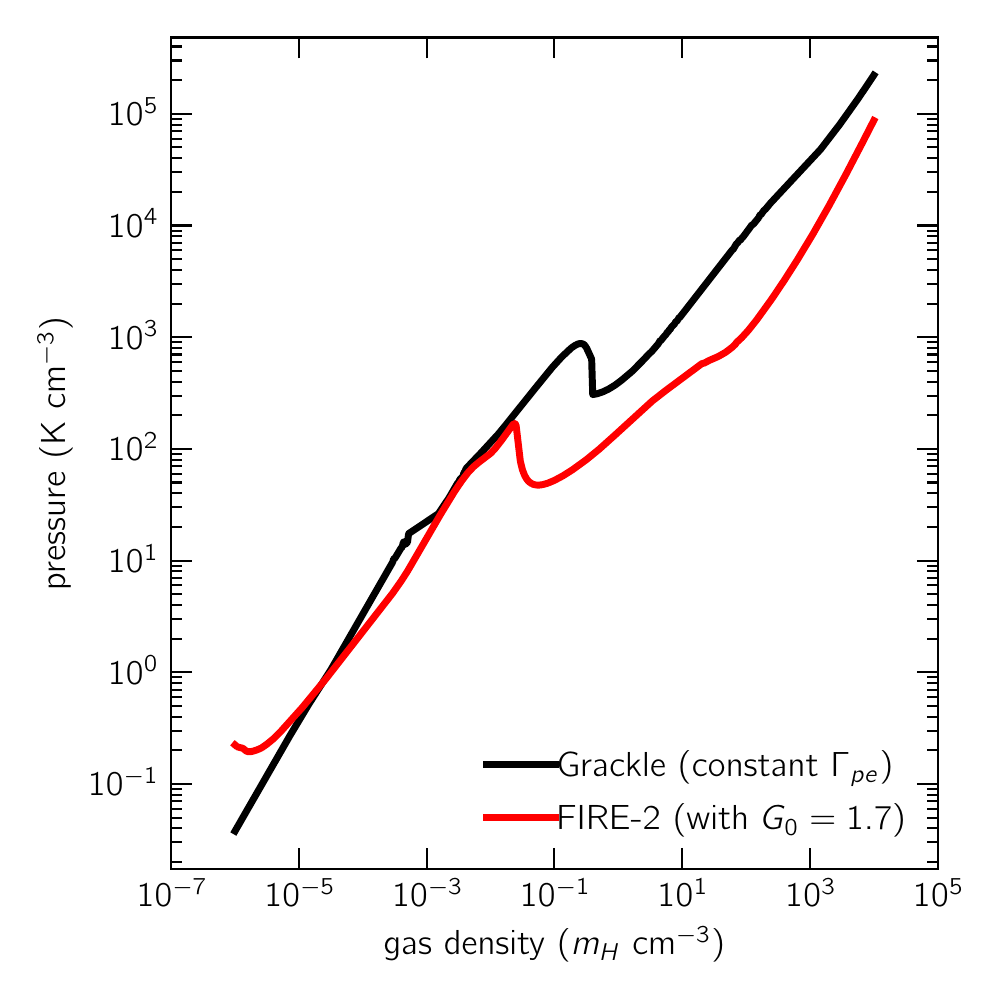}
    \caption{\emph{Left:} The temperature-density equilibrium cooling curves. \emph{Right:} The pressure-density equilibrium cooling curves.}
    \label{fig:cooling_curve_comparison}
\end{figure*}

In \autoref{fig:cooling_curve_comparison}, we show the equilibrium temperature and pressure as a function of density produced by the Grackle cooling code (version 2.2; \citealt{Smith_2017}), as used in this work, and the GIZMO cooling module as used in, e.g., the FIRE-2 simulations \citep{Hopkins_2018}, respectively.\footnote{The complete source code needed to reproduce these figures is publicly available in a GitHub repository: \url{https://github.com/BenWibking/cooling-curve-comparison}.} The unstable neutral medium is the phase for which $dP/dn<0$, i.e., where the slope is negative in the right panel of \autoref{fig:cooling_curve_comparison}; the stable warm and cold atomic phases correspond the regions with $dP/dn>0$ on the low- and high-density sides of this region, respectively. We see from the figure that the Grackle cooling curve features an unstable neutral phase between 943 K and 4276 K, corresponding to a gas density of 0.25 and 0.40 H cm$^{-3}$ and pressures of 306 K cm$^{-3}$ and 876 K cm$^{-3}$. The FIRE-2 cooling curve features an unstable neutral phase between 1092 K and 8980 K, between densities of 0.02 and 0.06 H cm$^{-3}$ and pressures of 47 and 168 K cm$^{-3}$ for solar neighbourhood FUV irradiation.

Based on \cite{Wolfire_2003}, we expect an unstable phase at pressures between 1960 K cm$^{-3}$ and 4810 K cm$^{-3}$ between densities of $0.86$ and $6.91$ H cm$^{-3}$ and temperatures of 258 K and 5040 K for solar neighbourhood ISM conditions. Clearly, neither cooling curve agrees quantitatively with expectations. However, while for Grackle the unstable pressure range is a factor of $\sim 5$ below that computed by \citet{Wolfire_2003}, for FIRE-2 the discrepancy is a factor of $30-50$. Indeed, with the FIRE-2 cooling code, the diffuse interstellar medium (largely shielded from ionisation due to young stellar populations) of galactic discs will be entirely composed of cold neutral medium, with no stable warm phase at all --- with the FIRE-2 cooling, such a phase exists only for pressures $P/k_B \lesssim 50$ K cm$^{-3}$, which is far below those pressures found in galactic discs and instead is more typical of the circumgalactic medium.

\begin{figure}
    \includegraphics[width=\columnwidth]{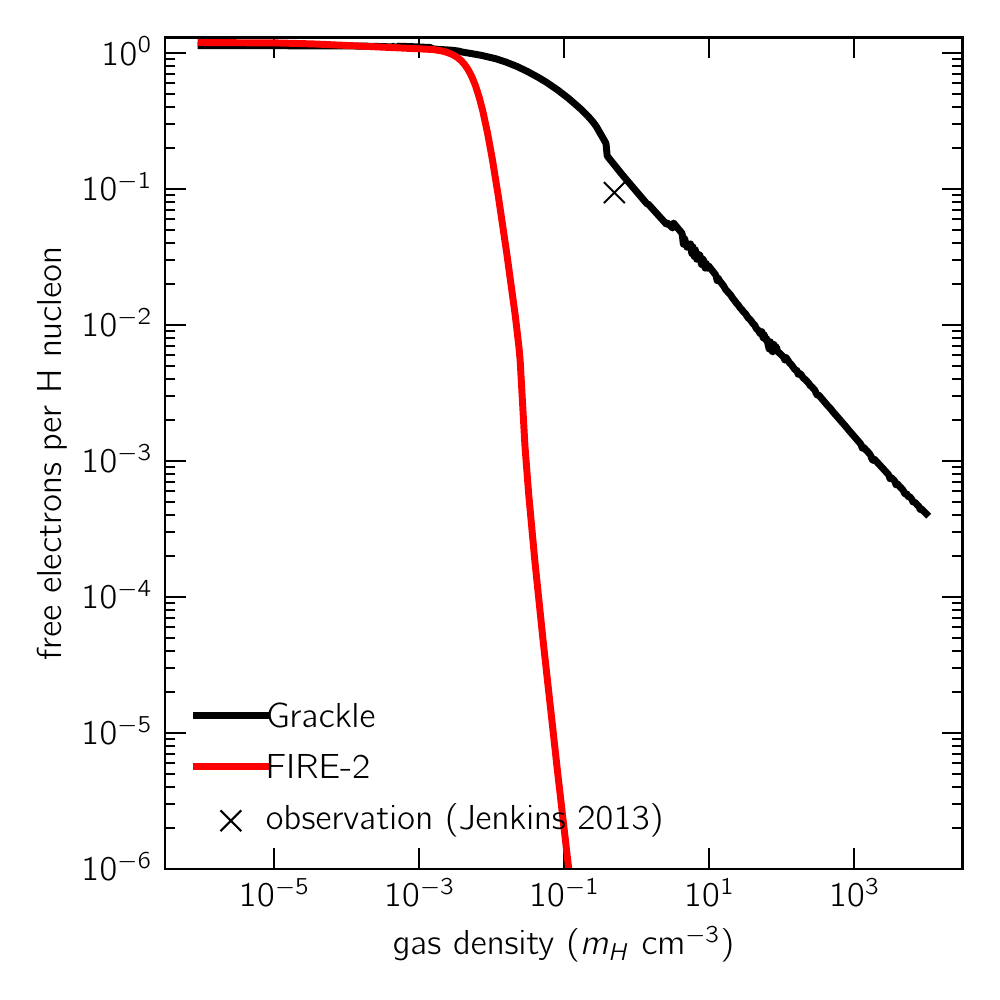}
    \caption{The number of free electrons per H nucleon in ionisation equilibrium as a function of gas density for both Grackle \citep{Smith_2017} and FIRE-2 \citep{Hopkins_2018} cooling.}
    \label{fig:ionisation_curve}
\end{figure}

The source of this discrepancy in the FIRE-2 cooling curve is the unphysically low grain photoelectric heating rate produced by the FIRE cooling module in the neutral atomic ISM. Specifically, the photoelectric heating efficiency [as defined by Eq. (20) of \cite{Wolfire_2003}] is 3--4 orders of magnitude too low when computed by GIZMO. This is a result of the model used for attenuating the ionising flux from the extragalactic UV background, which sharply cuts off the flux above gas densities of $\sim 0.0123$ cm$^{-3}$, resulting in an unphysically-low free electron density of $\sim 10^{-6}$ cm$^{-3}$ in the warm neutral ISM. In the non-public FIRE-2 radiative transfer code, there are additional local photoionising sources from young stellar populations, which partially mitigates this effect on the free electron number (but overestimates the photoionising flux due to the optically-thin approximation used for radiative transfer). However, the dominant source of ionisation in the neutral atomic ISM of the Galaxy is \emph{not} young stellar populations, but the combination of stellar EUV emission from old stellar populations (e.g., low-mass X-ray binaries) and the soft diffuse X-ray background produced primarily by X-ray line emission in supernova remnants \citep{Slavin_2000}, with an additional contribution from C$^{+}$ ionisation in regions of high FUV irradiation \citep{Wolfire_2003}. These sources are not included in the FIRE-2 ionisation model. Including these photoionising sources in models (e.g., \citealt{Wolfire_2003}) yields a free electron number density in the warm neutral interstellar medium consistent with the observationally-inferred free electron density in the solar neighbourhood of $\approx 0.047$ cm$^{-3}$ (assuming a hydrogen nucleon density $n_{\rm H} = 0.5$ cm$^{-3}$; section 8.1 of \citealt{Jenkins_2013}).

Grackle version 2.2 does not include a self-consistent attenuation model for UV background radiation, so we did not enable it in our simulations, and our Grackle cooling calculations are therefore in the optically-thin limit. As shown by the detailed radiative transfer calculations of \cite{Rahmati_2013} (c.f. Figure 3), optically-thin extragalactic ionisation remains a good approximation for calculating the ionisation state of interstellar gas up to densities of at least $\sim 10$ cm$^{-3}$, because the attenuation of the extragalactic background is very nearly compensated by an increase in ionising flux due to diffuse galactic ionising sources. The equilibrium number of free electrons per hydrogen nucleon as a function of gas density for both Grackle and FIRE is shown in \autoref{fig:ionisation_curve}. We note that the Grackle equilibrium ionisation predictions are broadly consistent with observations, whereas the FIRE equilibrium ionisation model is not.

Since the photoelectric heating efficiency scales as the free electron number density to the 0.73 power (Eq. 20 of \citealt{Wolfire_2003}), the ionisation state can therefore make a difference of many orders of magnitude in the photoelectric heating rate. Due to this effect, in the diffuse interstellar medium (which should be far from ionising radiation produced by young stellar populations), the FIRE-2 cooling model transitions between the warm and cold neutral phases at pressures and densities that are several orders of magnitude too low. In light of this qualitatively incorrect thermal structure, all conclusions regarding the thermal state of the interstellar medium in the FIRE-2 simulations (e.g., \citealt{Gurvich_2020}; \citealt{Pandya_2021}) should be critically re-examined.


\bsp	
\label{lastpage}
\end{document}